\documentclass{JINST}

\usepackage{natbib}
\usepackage{multirow}
\usepackage{subfigure}
\usepackage{aas_macros}

\title{X-ray polarimetry in Astrophysics with the Gas Pixel Detector}

\author{F. Muleri$^a$\thanks{Corresponding
author.}, R.~Bellazzini$^b$, A.~Brez$^b$, E.~Costa$^a$, F.~Lazzarotto$^a$, M.~Minuti$^b$, M.~Pinchera$^b$, A.~Rubini$^a$, P.~Soffitta$^a$ and G.~Spandre$^b$\\
\llap{$^a$}INAF/IASF Rome \\
  Via del Fosso del Cavaliere 100, I-00133 Rome, Italy \\
\llap{$^b$}INFN Pisa \\
  Largo B. Pontecorvo 3, I-56127 Pisa, Italy \\
  E-mail: \email{fabio.muleri@iasf-roma.inaf.it}}

\abstract{The Gas Pixel Detector, recently developed and continuously improved by Pisa INFN in
collaboration with IASF-Roma of INAF, can visualize the tracks produced within a low Z  gas by
photoelectrons of few keV. By reconstructing the impact point and the original direction of the
photoelectrons, the GPD can measure the linear polarization of X-rays, while preserving the
information on the absorption point, the energy and the time of individual photons. Applied to X-ray
Astrophysics, in the focus of grazing incidence telescopes, it can perform angular resolved
polarimetry with a huge improvement of sensitivity, when compared with the conventional techniques
of Bragg diffraction at 45$^\circ$ and Compton scattering around 90$^\circ$. This configuration is
the basis of  POLARIX and HXMT, two pathfinder missions, and is included in the baseline design of
IXO, the very large X-ray telescope under study by NASA, ESA and JAXA.}


\begin{document}

\section{Introduction}

Polarimetry is the last unexplored probe in High Energy Astrophysics, being particularly interesting in X-rays. In this energy domain the nonthermal emission is typically predominant and consequently a high degree of polarization is expected  \citep{Novick1975,Rees1975}. On the basis of a literature developed since the very beginning of X-ray astronomy, polarimetry can easily discriminate among competitive models \citep{Meszaros1988,Dyks2004,Bucciantini2005}. Moreover the relatively high fluxes potentially allow for collecting a sufficient number of photons to reach minimum detectable polarization at the level of a few percent and below. Indeed, even for an instrument without any systematic effect, a partial degree of polarization is always measured because of statistical fluctuations which can be reduced only with a larger amount of data.

Gas detectors able to image the track of photoelectrons are today the most valuable alternative to the instruments based on classical techniques (Bragg diffraction at 45$^\circ$ and Thomson/Compton scattering), which have succeeded only in the measurement of the polarization of one of the brightest object in the X-ray sky, the Crab Nebula \citep{Weisskopf1978,Dean2008}. In the following we describe the Gas Pixel Detector and briefly summarize the opportunity of missions currently discussed for this instrument.

\section{The Gas Pixel Detector}

The Gas Pixel Detector (GPD hereafter) has been developed by INFN of Pisa and IASF-Rome of INAF and
was the first device able to resolve the track of photoelectrons with a few keV energy, while
retaining a good quantum efficiency \citep{Costa2001,Bellazzini2006,Bellazzini2007}. In a similar
way other authors \citep{Colas2004} have developed a monolithic single-electron sensitive device by
coupling charge multipliers as Micromegas or Gas Electron Multipliers (GEMs) to the existing
Medipix2 CMOS sensor. The use of the Timepix chip, an evolution of Medipix2, also permits a 3D track
reconstruction by measuring the drift time or the total charge collected on a pixel in a kind of
micro-TPC detector \citep{Llopart2007}.

\begin{figure}
\centering
\subfigure[\label{fig:GPD_principle}]{\includegraphics[totalheight=5cm]{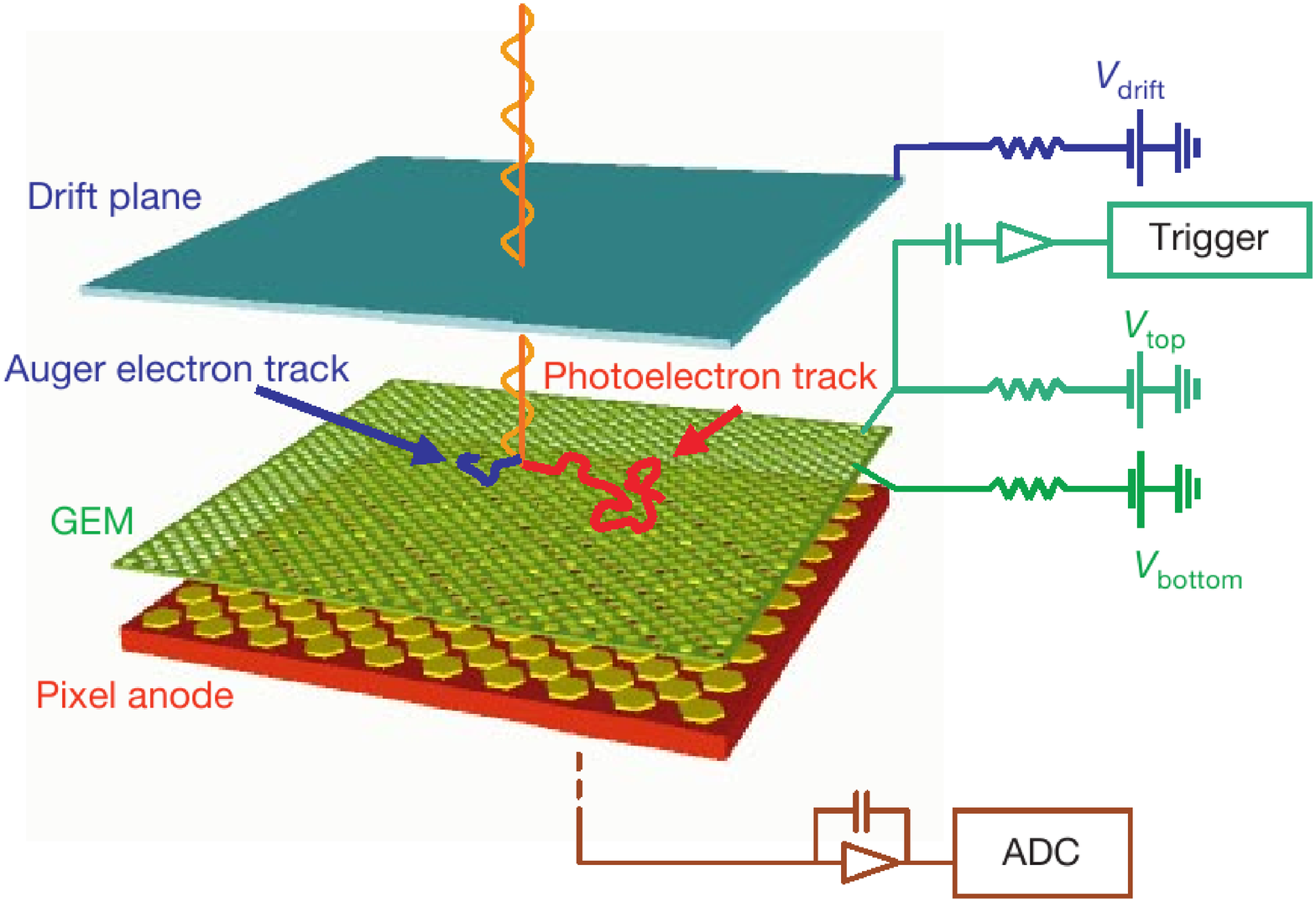}}\hspace{5mm}
\subfigure[\label{fig:Chip}]{\includegraphics[totalheight=5cm]{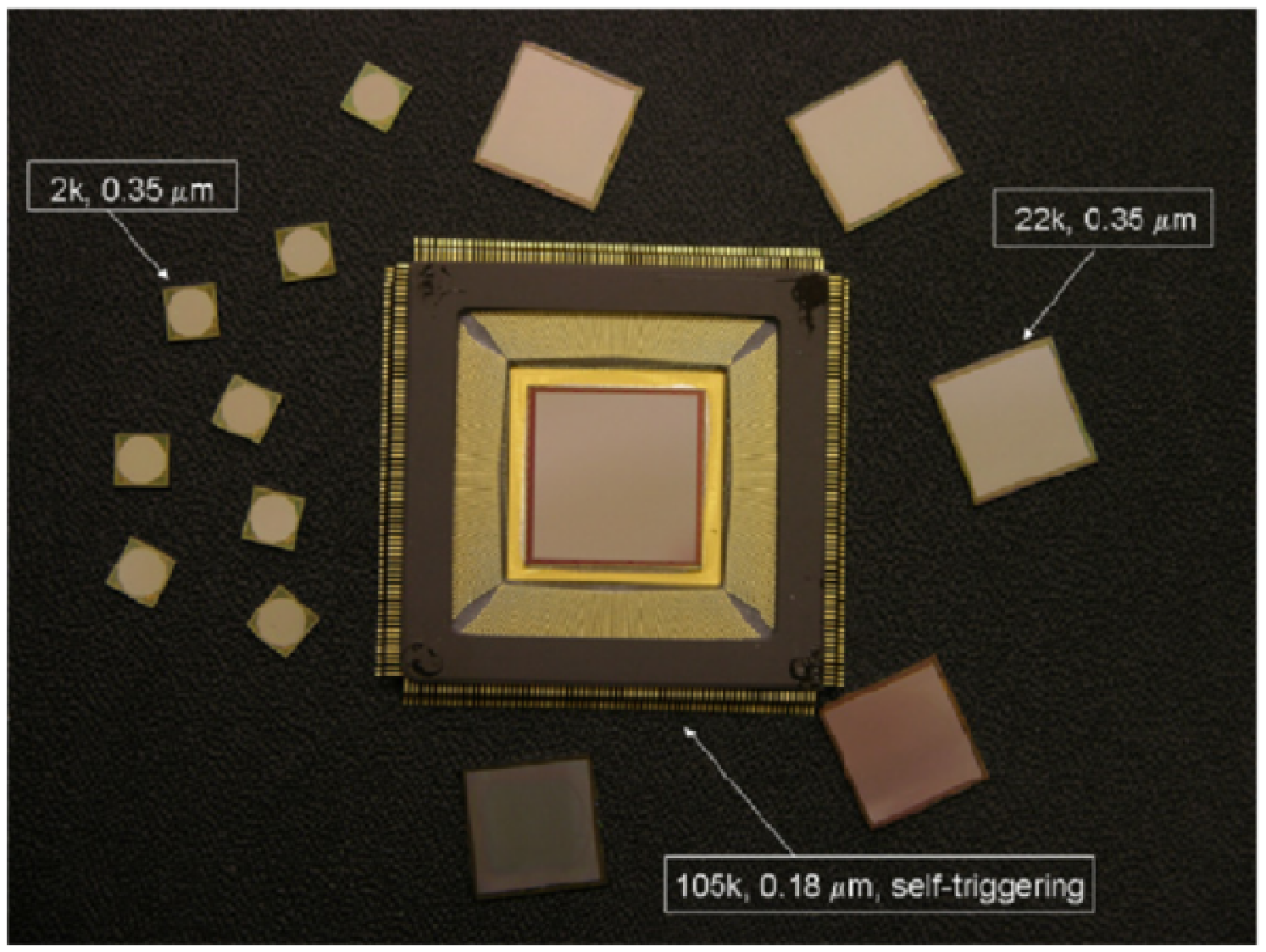}}
\caption{({\bf a}) Principle of operation of the GPD. ({\bf b}) The three generations of the ASIC chip developed by INFN of Pisa and INAF/IASF of Rome.}
\end{figure}

The presented GPD works basically like an array of standard yet exceptionally small proportional
counters with a common gas volume (see Fig.~\ref{fig:GPD_principle}). When an X-ray photon is
absorbed in a gas cell typically 1$\div$2 cm thick, a photoelectron is emitted. The charge produced
by ionization is collected and multiplied by a GEM and eventually read-out by a finely subdivided
custom VLSI ASIC, realized in 0.35~$\mu$m CMOS technology. The top metal layer of the CMOS is fully
pixellated to collect the charge produced in the gas volume and allow to get a true 2D imaging
capability, the actual breakthrough of the instrument (see Fig.~\ref{fig:Chip}). The main
characteristics of the GPD are reported in Table~\ref{tab:GPD}. The very small pixels (hexagonal
arranged at 50~$\mu$m pitch) are connected to full and independent electronics chains, which are
built immediately below the top layer of the ASIC and includes pre-amplifier, shaping amplifier,
sample and hold and multiplexer. The 105,600 pixels cover an area of 15$\times$15~mm$^2$ and are
divided into 16 clusters. Each cluster is further subdivided into mini-clusters of 4 pixels. A
trigger is generated when the charge collected by a mini-cluster is higher than a threshold
independently adjustable for each cluster, typically corresponding to a few electrons before the
amplification of the GEM. A rectangular region of 10 or 20 pixels (externally selectable) around the
mini-cluster(s) which triggered, called Region Of Interest (ROI), is fetched and read-out. This
self-triggering capability of the ASIC allows to read-out only about a thousand of pixels in place
of the whole matrix, reducing of a factor $\sim$100 the dead time. Immediately after the event, the
pedestals in the ROI are read-out (one or more times) and subtracted to data.

\begin{table}
\centering
\caption{Main characteristics of the current version of the GPD.}
\label{tab:GPD}
\begin{tabular}{rl}
Area: & 15$\times$15~mm$^2$ \\
Active area fill fraction: & 92\% \\
\hline
Window: & 50~$\mu m$, beryllium \\
Mixture: & He 20\% + DME 80\%, 1~atm \\
Cell thickness: & 1$\div$2~cm \\
\hline
GEM material: & gold-coated kapton \\
GEM pitch: & 50 $\mu m$ \\
GEM holes diameters: & 30 $\mu m$\\
GEM thickness & 50~$\mu m$ \\
Gain: & $\sim$500 \\
\hline
Pixels: & 300$\times$352, hexagonal pattern\\
Pixel noise: & 50 electrons ENC \\
Full-scale linear range: & 30000 electrons \\
\hline
Peaking time: & 3-10 $\mu s$, externally adjustable \\
Trigger mode: & internal, external or self-trigger \\
Self-trigger threshold: & 3000 electrons (10\% FS) \\
Pixel trigger mask: & individual \\
\hline
Read-out mode: & asynchronous or synchronous \\
Read-out clock: &up to 10~MHz \\
Frame rate: & up to 10~kHz  in self-trigger mode\\
Parallel analog output buffers: &1, 8 or 16 \\
\multirow{2}{*}{Access to pixel content:} & direct (single pixel) or serial (8-16 clusters, \\
&full matrix, region of interest) \\
\end{tabular}
\end{table}

\begin{figure}
\centering
\subfigure[\label{fig:GPD_sealed}]{\includegraphics[totalheight=6cm]{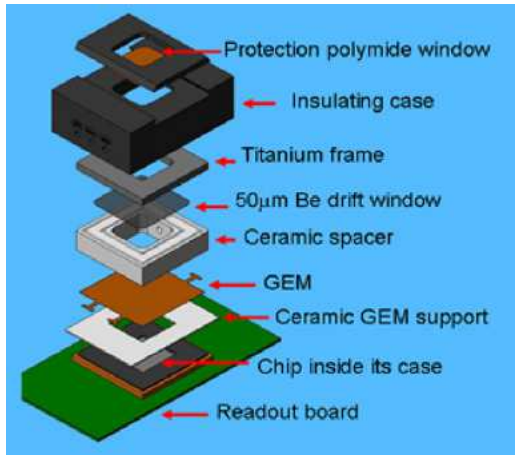}}\hspace{5mm}
\subfigure[\label{fig:GPD_box}]{\includegraphics[totalheight=6cm]{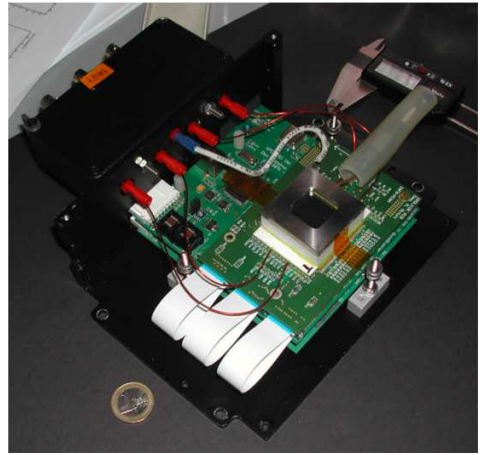}}
\caption{({\bf a}) Exploded view of the sealed GPD \citep{Bellazzini2007}. ({\bf b}) The box containing the detector, the front-end and processing electronics.}
\end{figure}

The gas cell of the detector, assembled in collaboration with Oxford Instruments Analytical Oy, is sealed even if it can be refilled to test different mixtures (see Fig.~\ref{fig:GPD_sealed}). The use of low-outgassing materials prevents the pollution of the mixture and no degradation of the performances of a prototype has been measured on time scale as long as years. Currently the operation of the instrument is optimized in the 1-10~keV energy range and the best sensitivity is achieved with an He-DME mixture at 1~atm. The whole instrument, including front-end and processing electronics to digitalize the analog signals, is contained in a box 140$\times$190$\times$70~mm$^3$ and connected with a USB to a standard PC running Windows~XP (see Fig.~\ref{fig:GPD_box}. The weight is less than 1.6~kg and the power consumption $\leq$5~W.

\begin{figure}
\centering
\subfigure[\label{fig:Track}]{\includegraphics[totalheight=5.5cm]{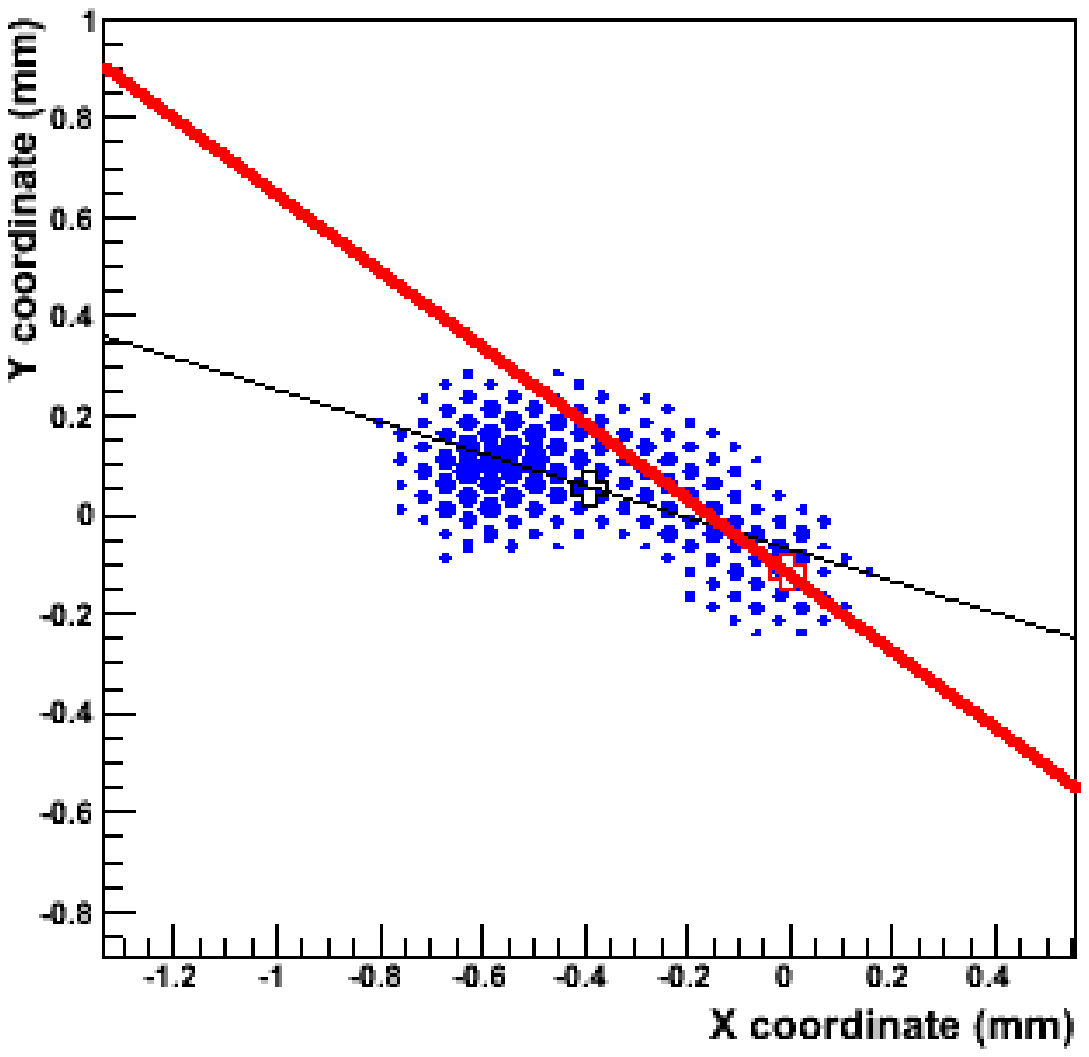}}\hspace{5mm}
\subfigure[\label{fig:Spectrum}]{\includegraphics[totalheight=5.5cm]{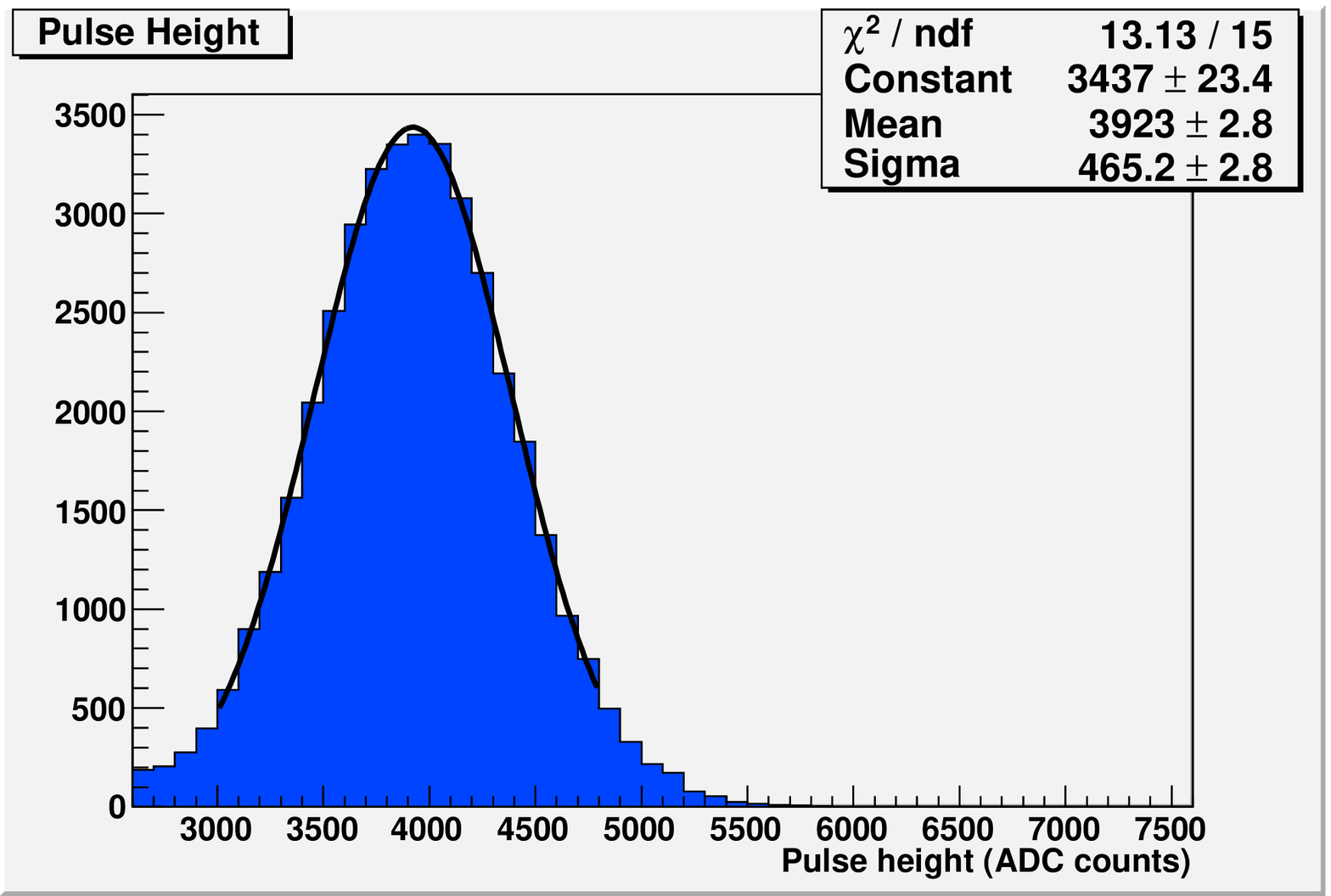}}
\caption{({\bf a}) Example of the reconstruction of a real track. The thick red line represents the better estimate of the photoelectron emission direction. ({\bf b}) Spectrum of 4.5~keV monochromatic photons. The energy resolution is $\leq$28\% at this energy.}
\end{figure}

An example of a real photoelectron track produced by 4.5~keV photons in a gas cell 1~cm thick and filled with He 30\% and DME 70\% at 1~atm is in Fig.~\ref{fig:Track}. The track is analyzed with an algorithm which reconstructs at best both the absorption point of the photons and the initial direction of emission (thick red line in Fig.~\ref{fig:Track}), despite the scatterings occurring with atomic nuclei which smear the track. The capability to measure the absorption point of the photons, with a resolution of the order of 150~$\mu$m, gives to GPD the unique possibility among gas polarimeters to image the source. The time of arrival and the energy of photons are also available, with the resolution of a good proportional counter. Currently these informations are acquired by the trigger of the acquisition and the total charge collected by the pixels, but the goal is to get them from the
signal of the GEM. This should assure a better timing ($\sim$10~$\mu$s) and spectral capabilities (20\% at 6~keV), even if current results are already encouraging. We measured an energy resolution $\lesssim$28\% at 4.5~keV for the current prototype (see Fig.~\ref{fig:Spectrum}). Scaling with energy, this results in an energy resolution of about 24\% at 6~keV.

\section{Astrophysical application}

\begin{figure}
\centering
\subfigure[\label{fig:ModCurve}]{\includegraphics[totalheight=5cm]{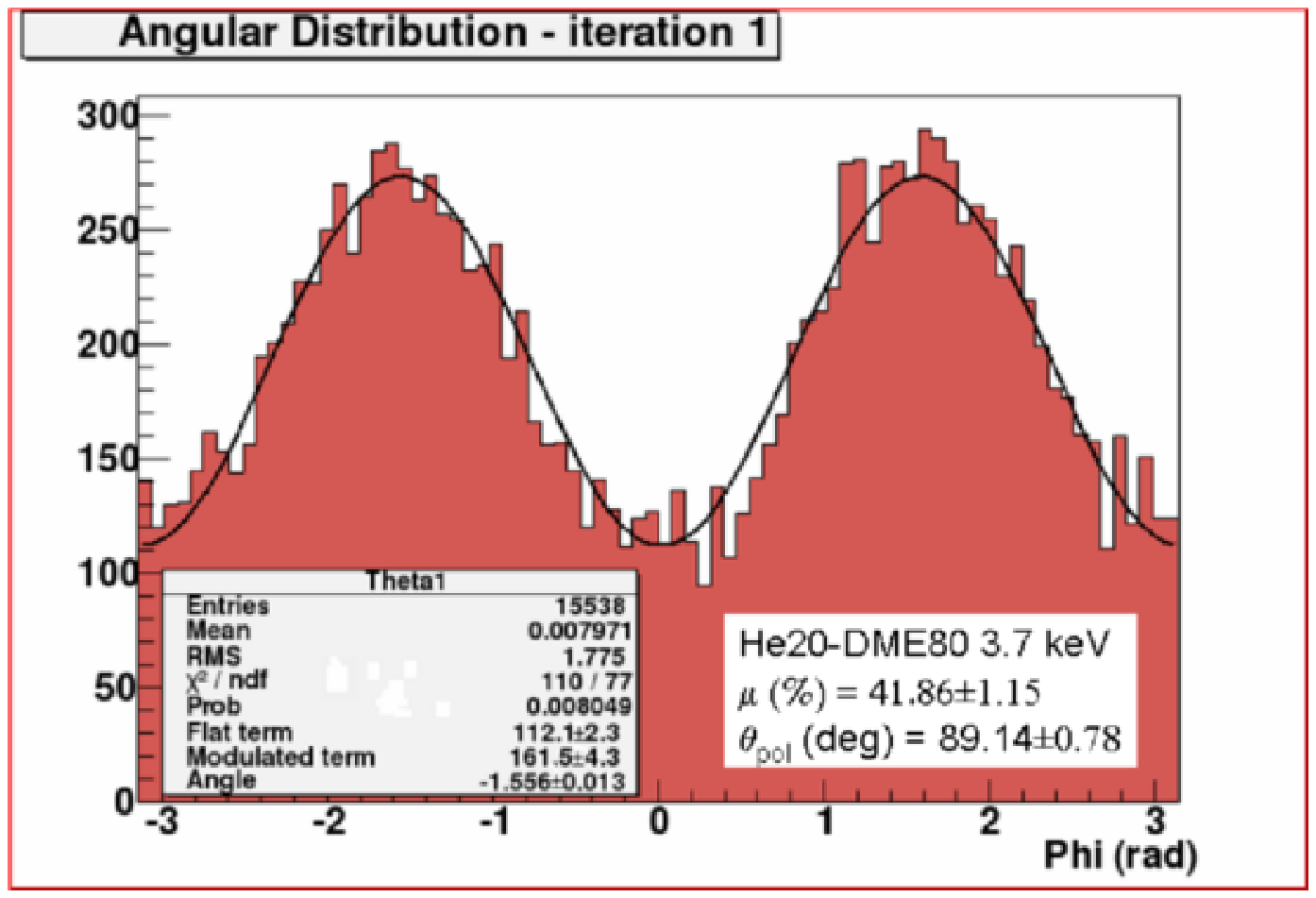}}\hspace{5mm}
\subfigure[\label{fig:ModFac}]{\includegraphics[totalheight=5cm]{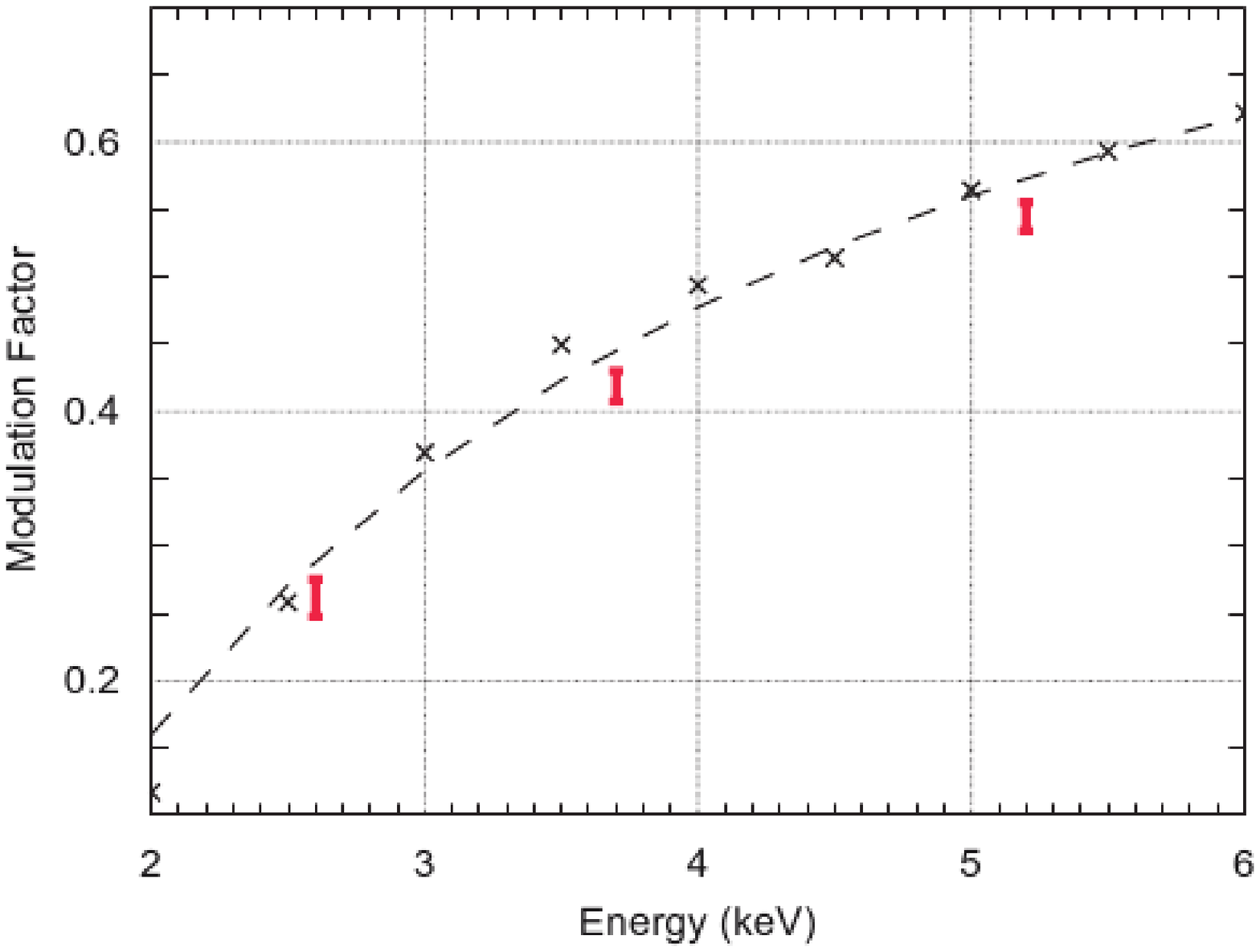}}
\caption{({\bf a}) Azimuthal distribution of photoelectrons for completely polarized photons at 3.7~keV, absorbed in a He 20\% and DME 80\% mixture. ({\bf b}) Comparison between measured modulation factor at 2.6, 3.7 and 5.2~keV and the expected values derived from Monte Carlo simulations \citep{Muleri2008}.}
\end{figure}

The capability of resolving photoelectrons tracks makes the GPD a sensitive polarimeter. The direction of emission brings memory of the polarization of the absorbed photons because the probability of emission in a certain direction is modulated with a term $\cos^2\phi$, where $\phi$ is the angle with the polarization vector. By constructing the histogram of the azimuthal directions of emission, the degree and angle of polarization are derived by the amplitude and the phase of the modulation respectively. In Fig.~\ref{fig:ModCurve} we report an example of the modulation measured at 3.7~keV for completely polarized photons. The amplitude of the modulation, called modulation factor $\mu$, is nearly 42\%.

Even if a further improvement is possible with other mixtures, the performance of the instrument has already reached a very good level with a mixture He 20\% and DME 80\% at 1~atm (gas cell 1~cm thick). In Fig.~\ref{fig:ModFac} there is the comparison between the measured modulation factor at 2.6, 3.7 and 5.2~keV and the expected value calculated with a Monte Carlo software which takes into account the physics of absorption, propagation and collection of the charges in the gas cell. On the basis of these results, the GPD can provide a sensitivity much higher than previous instruments, even in the context of a small and low cost mission (see below).

Another strong point of the GPD is its high readiness level. The existence of sealed detectors which
don't show any significant degradation after years of continuous operation makes us confident that
the pollution of the mixture will be very well under control on the time scale of a satellite
mission (a few years). A GPD was exposed at operative voltages at the Heavy Ions Medical Accelerator
in Chiba (HIMAC), Japan, to a flux of Fe ions corresponding to tens of years in LEO (Low Earth
Orbit). A prototype was subjected to vibration and thermovacuum tests (between -15$^\circ$C and
45$^\circ$ not in operation and at 10, 15, 20$^\circ$C in operation) which have proved the
capability to survive to launch and space environment. The possibility of destructive discharges
between GEM faces is rather low since the gain requested to be sensitive to the single primary
electron is only a few hundreds thanks to the very small pixel noise. Space-qualified processing
electronics, whose operation is analogous to the version already working in our laboratory, is under
construction and it will be able to acquire the analog signal from GEM to derive the energy and the
time of the event. The GPD doesn't need rotation because of the lack of any significant systematic
effect. For completely unpolarized photons produced with a Fe$^{55}$ radioactive source (lines at
5.9 and 6.5~keV) we measured a spurious modulation 0.18$\pm$0.14\% \citep{Bellazzini2010}.

The excellent polarimetric performances, together with the high readiness level, has allowed us to propose many missions with the GPD on-board. In each mission, the GPD is used together with an X-ray optics for a number of reasons. Beyond the obvious role of providing a sufficient collecting area, an X-ray optics can exploit at best the imaging capability of the instrument. This is important to resolve extended sources or confused regions, like the Galactic center. The presence of off-axis bright sources induces systematic effects which may mimic a spurious polarized signal. Another important advantage is that the background, a further source of systematic effects, is completely negligible in the focal plane of a telescope. The flux of the source under study has to be compared with the background, both internal and diffuse, only in the point spread function of the optics. This makes its contribution orders of magnitude lower than the fainter source accessible to polarimetry, which has to be relatively bright to collect enough photons in a reasonable observation time. Even for a short focal length optics (which implies a wider PSF and hence a higher background), the expected internal background is of the order of 4$\times$10$^{-6}$~c/s, while the contribution of the diffuse emission is $\sim$10$^{-8}$~c/s. The fainter source accessible has a flux >10$^{-3}$~c/s.

Two scenarios of possible missions have emerged in the last few years. The first is a small and low cost Italian mission, possibly dedicated to X-ray polarimetry, or an instrument to be carried on-board an international satellite; the second is the inclusion of the instrument among the focal plane instrumentation of a much larger international mission. These two possibilities differ for a number of reasons. A pathfinder mission could be launched in a few years, while a large satellite requires much longer developments and only a (small) fraction of time could be dedicated to X-ray polarimetry. Conversely, small missions are intended as all-purpose satellites to guide the development of more sensitive instruments and then suffer of many trade-offs, such as limited optics area, focal length and hence angular resolution.

\begin{figure}
\centering
\subfigure[\label{fig:PolariX}]{\includegraphics[totalheight=5.3cm]{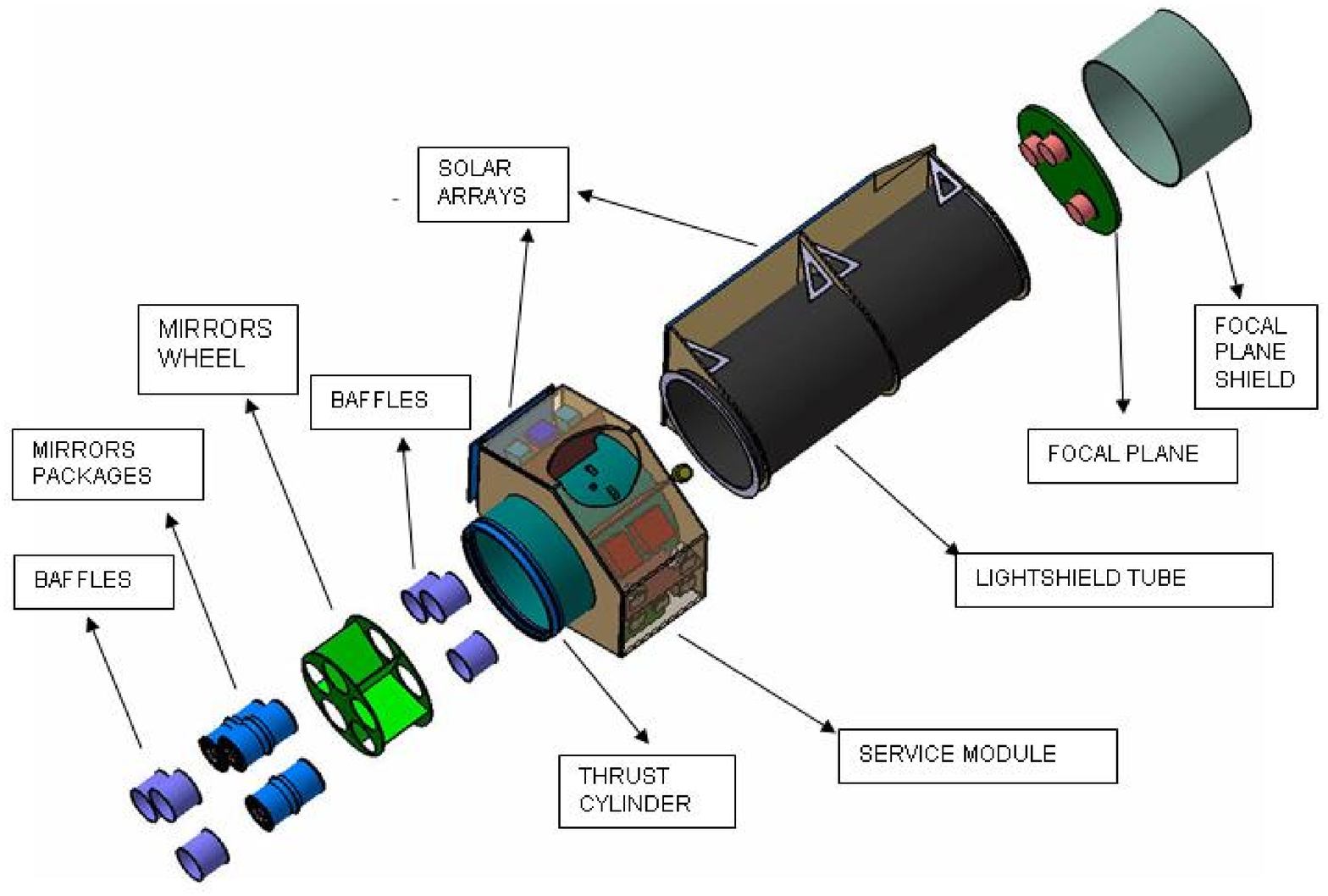}}\hspace{5mm}
\subfigure[\label{fig:FocalPlane}]{\includegraphics[totalheight=5.3cm]{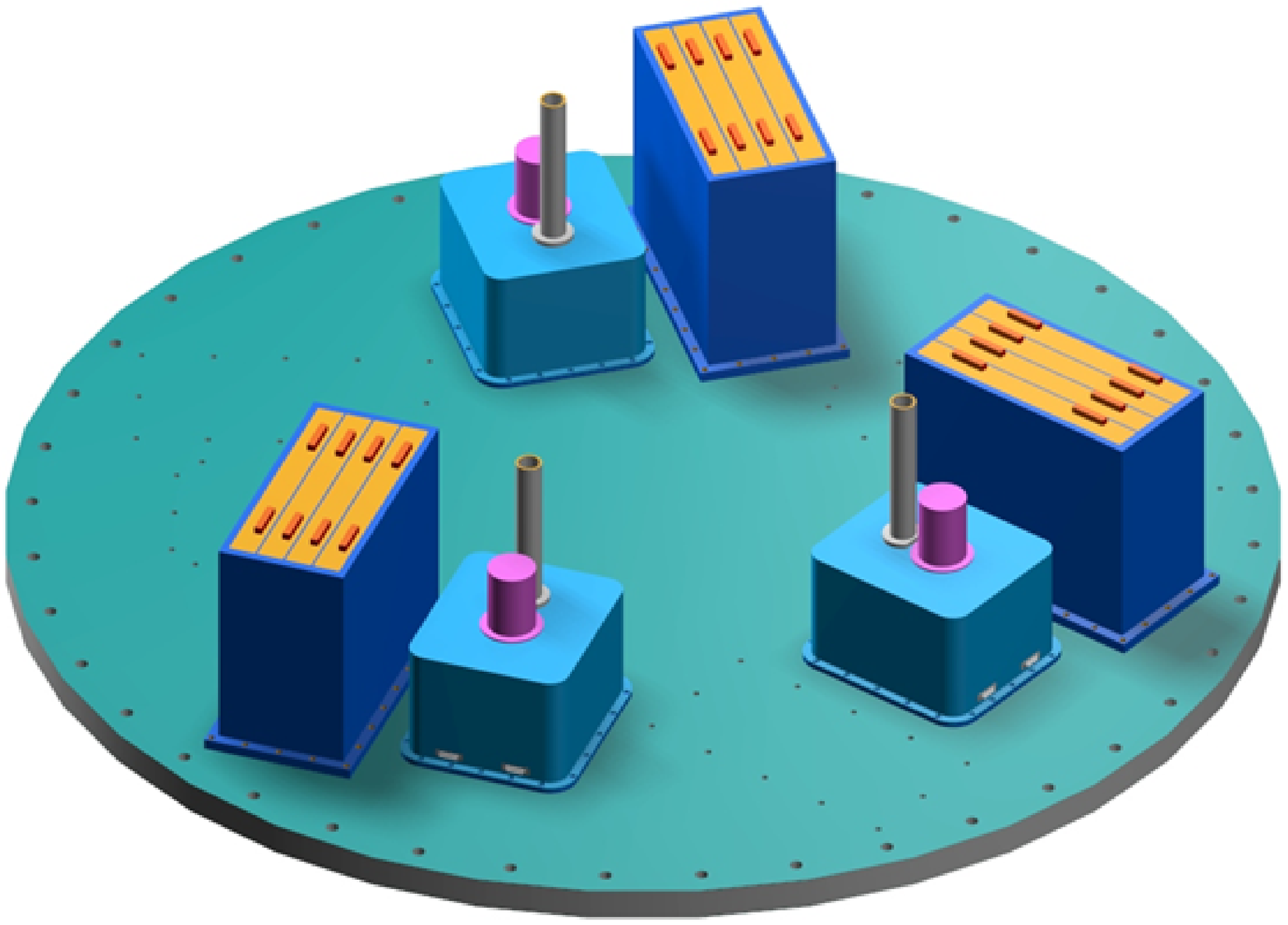}}
\caption{({\bf a}) Design of POLARIX. The bus can host three telescopes, already built for the Jet-X
mission, plus two further units if additional funds are available. ({\bf b}) Focal plane layout of
POLARIX.}
\end{figure}

Within the context of small missions, two opportunities have passed a certain degree of selection.
The first is POLARIX, a completely Italian mission dedicated to X-ray polarimetry with the GPD
\citep{Costa2006,Costa2009}, waiting for the possible selection to the launch after that the phase~A
study ended in December 2008. Three detectors, possibly filled with different mixtures to be
optimized in slightly different energy bands, are placed in the foci of as many telescopes (with a
focal length of 3.5~m), already built for the \emph{Jet-X} instrument on-board \emph{Spectrum X-ray
Gamma} which unfortunately has never flown (see Fig.~\ref{fig:PolariX}). In the case of the GPD,
more telescopes provide comparable sensitivity of a single optics of equal area since the background
is always negligible. However, the use of identical units can strongly reduce costs because, after
the construction of mandrels for a small number of different shells for the first telescope, the
replica of identical mirrors is a relatively inexpensive procedure. For this reasons, the bus can
host two further telescope units if additional fundings are available.

The focal plane layout is in Fig.~\ref{fig:FocalPlane}. On the side of each detector, there are the front-end electronics and the high voltage power supply. A filter wheel is also present to place in front of the detector calibration sources, a filter to reduce the flux of exceptionally bright sources and a diaphragm to exclude sources in the field of view much brighter than that observed. A single control electronics is devoted to manage the focal plane instruments and forward data of all detectors to the bus. In case of very bright sources, the control electronics is also in charge to perform tracks reconstruction on-board. The transmission of the whole informations, namely the charge collected by each pixel, the energy and the time of the event, may be too cumbersome, even after the suppression of the pixels in the ROI which didn't collected any charge (zero suppression). For this reason, an option is to transmit only the essential data, namely the energy, the time of the event, the absorption point, the reconstructed direction of emission plus some quality parameter of the track. An alternative solution is to save all data of strong sources in a large on-board memory and transmit them during the subsequent observation of faint sources. This problem is not critical for pathfinders, but for the large mission scenario the on-board analysis will be the standard procedure for sources above a few hundreds of mCrab because of the much higher counting rate.

The second opportunity of small mission is the inclusion of the GPD on-board the Chinese mission Hard X-ray Modulation Telescope, whose primary scientific object is a sensitive survey in the hard X-rays. However a half of the observation time will be dedicated to pointed observations and two GPDs, with as many dedicated telescopes, could be inserted in the scientific payload of HXMT to perform contemporarily polarimetry and measurements in the hard X-ray range. Indeed these two observations are deeply connected, both being the signature of nonthermal processes. The Italian Space Agency is currently negotiating with the Chinese counterpart for the Italian contribution to HXMT mission.

\begin{figure}
\centering
\subfigure[\label{fig:CrabNebula}]{\includegraphics[totalheight=5.3cm]{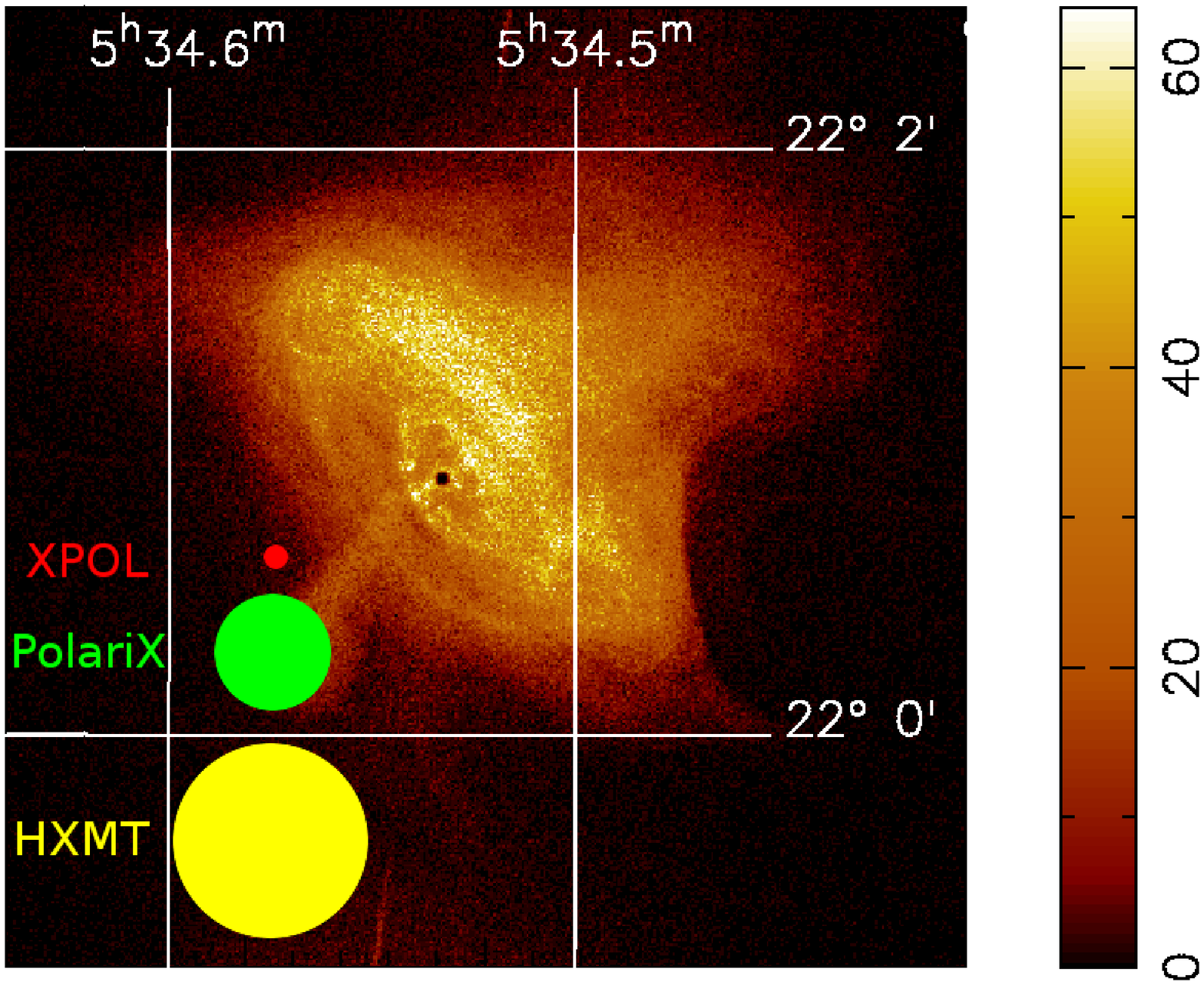}}\hspace{5mm}
\subfigure[\label{fig:MDP_PolariX}]{\includegraphics[angle=90,totalheight=5.3cm]{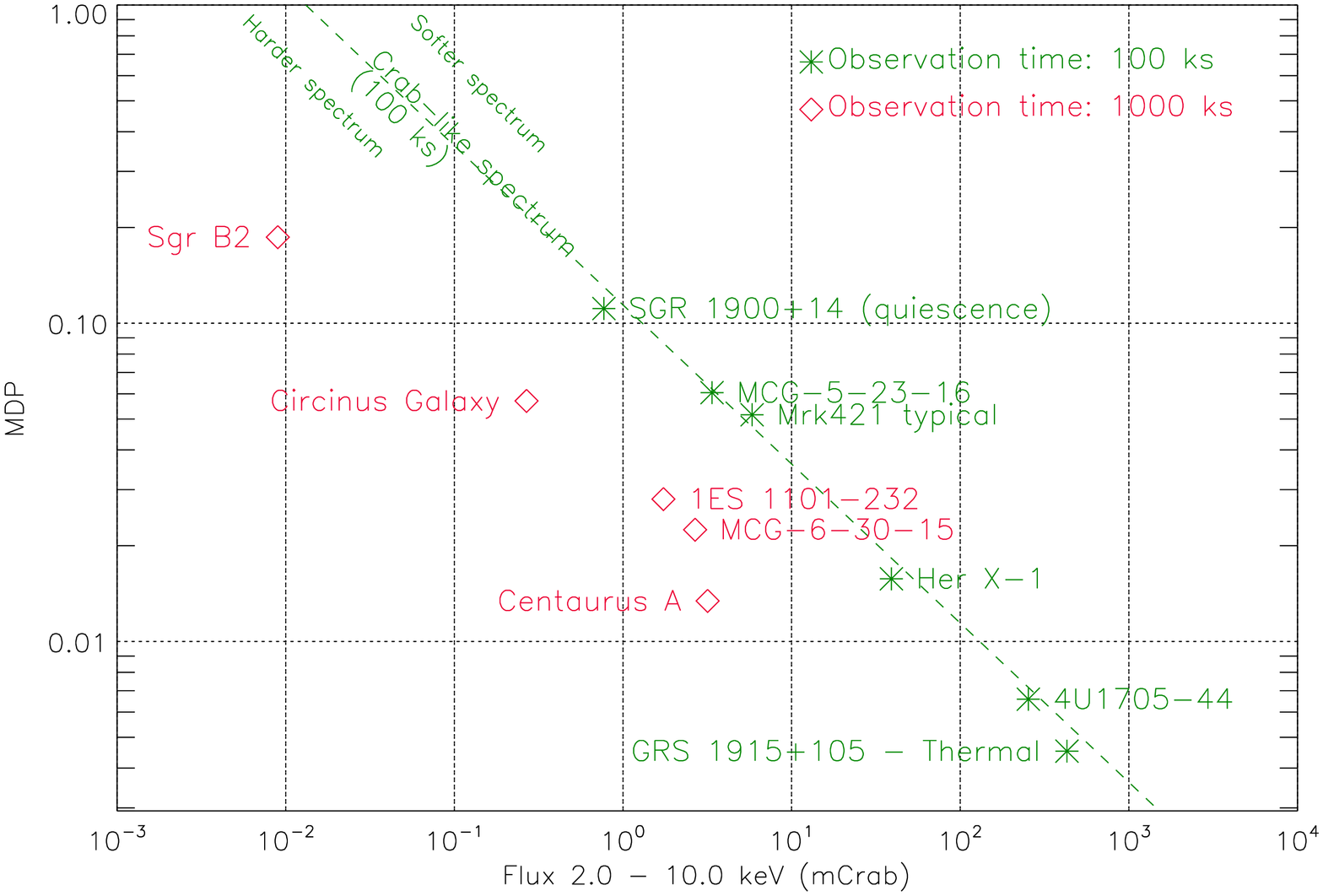}}
\caption{({\bf a}) Comparison of the angular resolution of small missions and XPOL on-board IXO.
({\bf b}) Minimum detectable polarization as a function of flux for POLARIX.}
\end{figure}

The focal plane of the polarimeter on-board HXMT shares many similarities with POLARIX. The main
difference between these two small missions is the different telescope. The Jet-X optics are too
heavy and can't fit the volume in the fairing of the Long March launcher of HXMT. Then a dedicated
telescope with a shorter focal length (f=2.1~m) has to be built. Producing thin shells (between 100
and 200~$\mu$m), modern technologies allow to save weight at the cost of a controlled degradation of
angular resolution. In Fig.~\ref{fig:CrabNebula} we compare the angular resolution of POLARIX
(24~arcsec) and HXMT (40~arcsec). The shorter focal length also makes the area of the HXMT telescope
softer, but thanks to the more modern coating (iridium and carbon) the area is larger at ~3~keV
(614~cm$^2$ vs 341~cm$^2$), where the response of the GPD peaks.

In Fig.~\ref{fig:MDP_PolariX} the minimum detectable polarization (at the 99\% confident level) for
POLARIX, chosen as a benchmark between small missions, is reported. We can reach 1\% for a 100~mCrab
source in $\sim$1~day. Observations of even fainter sources can be planned with longer observations
(up to 10~days). Measurements with systematic effects below 1\% are well within the possibilities of
the GPD since its intrinsic axial symmetry (we measured a spurious modulation 0.18$\pm$0.14\%
for completely unpolarized photons from Fe$^{55}$ \citep{Bellazzini2010}).

The instrument XPOL (X-ray Polarimeter), based on the Gas Pixel Detector and on the same design as small missions, is also inserted in the focal plane of the multi-purpose \emph{Internation X-ray Observatory}, a joint effort of NASA, ESA, and JAXA, whose launch is scheduled for 2021. The focal length of 20~m, achieved by an extendable bench, and the huge area ($\sim$1.5~m$^2$ at 3~keV) makes it a great step forward from pathfinder missions. The current chip is almost ready for the use on-board IXO, the only improvement being the reduction of the dead time to sustain the high counting rate expected (6600~c/s for 1~Crab source). The read-out clock will be speeded-up and the margin around the triggered pixel will be reduced to shrink the ROI and avoid a large number of zeros. Even if the observation time dedicated to polarimetry will be $\sim$10\% of the total, faint sources (1~mCrab) will be accessible at the level of 1\% for 1~day of observation, i.e. XPOL will be 100 times more sensitive than pathfinders. Another unique capability is the fine angular resolution (6~arcsec) which allows to pinpoint all the principal structures of extended sources, like the Crab Nebula (see Fig.~\ref{fig:CrabNebula}), or extragalatic jets, which must be resolved from the close nuclear emission.

\section{Conclusion}

The Gas Pixel Detector is one of the most advanced instruments to image the tracks of photoelectrons
in a gas, both for performances and readiness level. It allows to measure the linear polarization in
the energy range $\sim$1-10 keV, reconstructing also the impact point, the energy and the time of
the events. The unique possibility to join the polarimetric informations to timing, spectral and
imaging capabilities is particularly interesting in X-ray Astrophysics. A large literature,
unfortunately still without a significant experimental feedback, describes polarimetry as a
fundamental tool to distinguish among competitive models, often equivalent on the basis of spectral
or timing informations alone. The GPD provides very concrete possibilities to perform measurements
at the level of 1\% and below on board small missions, like POLARIX or HXMT, to be launched in a few
years. In the future, the GPD will be also part of the International X-ray Observatory which will
definitely elevate X-ray polarimetry as a systematic tool for a deeper understanding of many
different classes of astrophysical sources.

\acknowledgments

\bibliography{References}
\bibliographystyle{unsrt}

\end{document}